\documentclass[compsoc, conference, letterpaper, 10pt]{IEEEtran}
\usepackage{array}
\usepackage{verbatim}
\usepackage{multirow}
\usepackage{graphicx}
\usepackage[unicode=true]{hyperref}
\usepackage{multicol}
\usepackage[export]{adjustbox}
\usepackage{enumitem}
\usepackage{wrapfig}
\usepackage{amsmath}
\usepackage{pifont}
\usepackage{makecell}
\usepackage{lipsum}
\usepackage{pgfplots}
\pgfplotsset{compat=1.14}
\usepackage{pgfplotstable}
\usepackage{color}
\usepackage[colorinlistoftodos]{todonotes}
\usepackage[cm]{sfmath}
\usepackage[ruled]{algorithm2e}
\usepackage{url}
\usepackage[all]{background}

\graphicspath{ {images/} }

\ifCLASSOPTIONcompsoc
  \usepackage[nocompress]{cite}
\else
  \usepackage{cite}
\fi

\ifCLASSOPTIONcompsoc
	\usepackage[caption=false,font=footnotesize,labelfont=sf,textfont=sf]{subfig}
    \else
    \usepackage[caption=false,font=footnotesize]{subfig}
\fi

\begin{document}

\SetBgContents{IEEE Workshop on Security \& Privacy on the Blockchain, 2018}
\SetBgPosition{-0.8cm,-6.5cm}
\SetBgOpacity{0.7}
\SetBgAngle{90.0}
\SetBgScale{1.3}

\title{Incentivized Delivery Network of IoT Software Updates \\ Based on Trustless Proof-of-Distribution}

\author{\IEEEauthorblockN{Oded Leiba, Yechiav Yitzchak, Ron Bitton, Asaf Nadler, Asaf Shabtai}
\IEEEauthorblockA{Department of Software and Information Systems Engineering\\
Ben-Gurion University of the Negev\\
Beer-Sheva, 8410501, Israel}
\{odedlei,yitzchak,ronbit,asafnadl\}@post.bgu.ac.il, shabtaia@bgu.ac.il
}

\maketitle

\begin{abstract}
The Internet of Things (IoT) network of connected devices currently contains more than 11 billion devices and is estimated to double in size within the next four years. 
The prevalence of these devices makes them an ideal target for attackers. 
To reduce the risk of attacks vendors routinely deliver security updates (patches) for their devices. 
The delivery of security updates becomes challenging due to the issue of scalability as the number of devices may grow much quicker than vendors' distribution systems. 
Previous studies have suggested a permissionless and decentralized blockchain-based network in which nodes can host and deliver security updates, thus the addition of new nodes scales out the network. 
However, these studies do not provide an incentive for nodes to join the network, making it unlikely for nodes to freely contribute their hosting space, bandwidth, and computation resources.

In this paper, we propose a novel decentralized IoT software update delivery network in which participating nodes (referred to as \textit{distributors}) are compensated by vendors with digital currency for delivering updates to devices. 
Upon the release of a new security update, a vendor will make a commitment to provide digital currency to distributors that deliver the update; the commitment will be made with the use of smart contracts, and hence will be public, binding, and irreversible. 
The smart contract promises compensation to any distributor that provides \emph{proof-of-distribution}, which is unforgeable proof that a single update was delivered to a single device. 
A distributor acquires the proof-of-distribution by exchanging a security update for a device signature using the Zero-Knowledge Contingent Payment (ZKCP) trustless data exchange protocol. 
Eliminating the need for trust between the security update distributor and the security consumer (IoT device) by providing fair compensation, can significantly increase the number of distributors, thus facilitating rapid scale out.

\end{abstract}

\section{\label{sec:introduction}Introduction}
The number of IoT devices is continuously increasing. 
According to Gartner Inc.\footnote{\url{https://www.gartner.com/newsroom/id/3598917}} 11 billion connected "things" will be in use in 2018, and this figure will reach up to 20 billion by the end of 2020.
The massive growth in the number of connected IoT devices, and particularly their essential need for security software and/or firmware upgrading (patching), has resulted in the search for reliable and trusted solutions for distributing software updates on a large scale. 

Traditional software update services mainly rely on a client-server architecture, where bandwidth consumption and maintenance issues impose high costs and put security and availability at risk. 
IoT vendors that wish to provide software update services for their products are required to maintain large-scale data centers to support software distribution to millions of devices\footnote{\url{http://www.channelpartnersonline.com/2017/12/05/aws-dell-emc-among-vendors-in-rapidly-growing-iot-services-market/}}.
The wide range of IoT devices (and firmware) also poses a challenge to the software update services commonly used for managing updates within organizations.
Consequently, vendors delegate the task of updating software to the end users who are not always aware of the security issues and the importance of keeping the devices updated.

For these reasons many IoT devices are not consistently being updated and remain vulnerable to known threats \cite{bekara2014security}, \cite{ray2017patching}, \cite{kumar2016security}. 
Previous works have targeted the availability and scalability challenges by using a permissionless blockchain-based decentralized network instead of a private centralized vendor network \cite{lee2017blockchain},\cite{boudguiga2017towards}. 
However, these solutions provide no incentive for non-vendor nodes to host and deliver security updates, and may face limitations similar to those faced by centralized vendor networks. 

In this paper we propose a novel \emph{decentralized} and \emph{incentivized} IoT update delivery network based on trustless \emph{proof-of-distribution}.
The proposed framework utilizes blockchain, smart contracts, zero-knowledge contingent payment (ZKCP), and a decentralized storage network.
Within this framework participating nodes (i.e., distributors) are compensated for delivering software updates to IoT devices.
More specifically, when a new security update is released, a vendor will agree to compensate (with digital currency) distributors who deliver the update.
The agreement is bound by a smart contract, making it public and irreversible, which guarantees compensation to any distributor that provides proof-of-distribution, i.e., unforgeable proof that a single update has been delivered to a single IoT device. 
A distributor acquires the proof-of-distribution by exchanging a security update for a device signature.
To ensure trustless data exchange we use the Zero-Knowledge Contingent Payment protocol. 
Eliminating the need for trust between the software update distributor and the consumer (IoT device) by providing fair compensation, can encourage competition, significantly increase the number of distributors and overall efficiency, thus allowing rapid scale out.

The main contributions of the proposed framework are:\\
\textbf{Availability.} The proposed framework ensures high network availability compared to the current client-server architecture.\\
\textbf{Programmable incentivization.} The framework makes it straightforward to support different prioritizations through minor adjustments to the smart contract. Vendors can decide to push forward specific critical security updates, prioritize specific IoT clients, or other relevant settings.\\
\textbf{Auditability.} The proposed framework allows vendors to monitor and track the software download of its devices.\\
\textbf{Integrity.} The proposed framework provides high integrity of the software.

The rest of the paper is organized as follows.
Section \ref{sec:background} provides a brief introduction to the building blocks of our proposed ecosystem: blockchain, smart contract, and zero-knowledge proofs.
Section \ref{sec:relatedWork} summarizes previous works in this domain, highlighting the manner in which the proposed ecosystem addresses the limitations of prior work.
The proposed ecosystem is described in Section \ref{sec:framework}.
In Section \ref{sec:ta} we briefly discuss potential attacks and their handling, and provide a formal claim and proof of the fair exchange. In Section \ref{sec:discussion} we discuss the limitations of the proposed framework and finally, in Section \ref{sec:conclusion-future-work} we conclude with the paper's contributions and introduce possible directions for future work.

\section{\label{sec:background}Background}
In this section we introduce the technologies utilized in the proposed framework, providing the necessary background to understand the protocol and its analysis.

\subsection{\label{sec:background-blockchain}Blockchain}
Blockchain was introduced alongside Bitcoin in Satoshi Nakamoto's white paper~\cite{nakamoto2008bitcoin}. 
A blockchain is a data structure that can be regarded as a public ledger where groups of messages are stacked one on top of the other.
These groups of messages are named "blocks" and with the use of digital signatures and a distributed consensus algorithm, users in a non-synchronous configuration (i.e., do not necessarily agree on time and order of messages) can irreversibly agree with high probability on a specific order of blocks.

The agreement on the order of blocks is useful for the prevention of currency "double-spending" (where messages within the blocks correspond to money transactions) as the latter spending can be mutually eliminated by the users, and is therefore the foundation for Bitcoin and various other cryptocurrencies~\cite{buterin2014next,eyal2016bitcoin, pass2017fruitchains}.

Concretely, each transaction message within a block contains the coin transfer value, the redeem terms, an input transaction and the signature of the transaction's author for authenticity. 
The redeeming terms can be referred to as a boolean predicate such that only a true evaluation can make use of the transaction.
Because published transactions are a part of a block, they are irreversible and thus redeeming terms can serve to vouch for coin in exchange for a truth assignment, which is the foundation of "smart contracts".

\subsection{\label{sec:background-smartContracts}Smart Contracts}
A smart contract~\cite{szabo1997idea} is a protocol that enforces the negotiation or performance of an agreement. 
In the context of blockchain-based cryptocurrencies, it is realized by a traceable and irreversible transaction that can be redeemed only if a set of terms are met.
At their essence, smart contracts don't differ from any other kind of transaction and are primarily named likewise when used with redeeming terms that are more complex than verifying the receiver's identity. 
However, the contextual notion of being addressed to anyone who qualifies to specific terms instead of a specific address makes smart contracts valuable to nodes who wish to satisfy these terms and collect the value.

The features of blockchain-based smart contracts evolved from the Bitcoin scripting language that is a simple, stack-based, and purposefully non-Turning-complete, to a new blockchain paradigm initiated with Ethereum~\cite{buterin2014next} which offers Turing-complete stateful languages for writing smart contracts.
The most widely adopted example of such language used for developing smart contracts in Ethereum is Solidity which is a JavaScript-like language.
Rootstock~\footnote{\url{http://www.the-blockchain.com/docs/Rootstock-WhitePaper-Overview.pdf}} is another example of a blockchain platform offering equivalent smart contract capabilities.
Hereafter, the use of the term smart contracts in this paper refers to the Ethereum implementation.

\subsection{\label{sec:background-zksnark}zk-SNARKs}
Zero-Knowledge Succinct Non-Interactive ARguments of Knowledge (zk-SNARKs)\cite{gennaro2013quadratic, bitansky2013succinct, ben2013snarks, ben2014succinct} is an efficient and secure system for proving and verifying zero-knowledge proofs.
It allows a \textit{prover} to (efficiently) convince a \textit{verifier} that she possesses knowledge of a secret parameter, called a \textit{witness}, satisfying certain properties, without revealing anything about the secret to the verifier or anyone else. For example, a prover can convince a verifier that it has knowledge of a secret $w$ which is the hash preimage of some value $x$, without revealing anything about $w$.
Because zk-SNARKs are \textit{succinct}, proofs are very short and easy to verify.
More formally, let $L$ be an \textsf{NP} language and $C$ be a decision circuit for $L$.
A trusted party conducts a one-time setup phase that results in two public keys: a proving key $pk$ and a verification key $vk$.
The proving key $pk$ allows any (untrusted) prover to generate a proof $\pi$ attesting that $x \in L$ for an instance $x$ of her choice. 
The non-interactive proof $\pi$ is both \textit{zero-knowledge} and \textit{proof-of-knowledge}. 
The proof $\pi$ has a constant size and can be verified in time that is linear in $|x|$.

A zk-SNARK for circuit satisfiability consists of the following three polynomial-time algorithms:
\begin{itemize}
\item $Gen(1^{\lambda},C) \rightarrow (pk, vk)$. On security parameter $\lambda$ and a decision circuit $C$, $Gen$ probabilistically samples $pk$ and $vk$. Both keys are published as public parameters and can be used to prove/verify membership in $L_c$.
\item $Prove(pk, x, w) \rightarrow \pi$. 
On input proving key $pk$, instance $x$ and witness for the \textsf{NP}-statement $w$, the \textit{prover} $Prove$ outputs a non-interactive proof $\pi$ for the statement $x \in L_c$.
\item  $Verify(vk, x, \pi) \rightarrow \{0,1\}$.
On input verifying key $vk$, an instance $x$, and a proof $\pi$, the \textit{verifier} $Verify$ outputs 1 if $x \in L_c$.
\end{itemize}

\subsection{\label{sec:background-zpf}Zero-Knowledge Contingent Payments (ZKCP)}
Zero-Knowledge Contingent Payment (ZKCP)~\cite{wiki2011zero} is a two-party protocol allowing the fair exchange of information for payment (e.g., cryptocurrency coins) in a setting where the two parties do not trust each other to deliver. 
Besides being private and secure in theory, it was practically used in Bitcoin with the use of zk-SNARKs~\footnote{https://bitcoincore.org/en/2016/02/26/zero-knowledge-contingent-payments-announcement/}.

\section{\label{sec:relatedWork}Related Works}
Our work falls within the following domains: IoT devices' patching and decentralized storage networks (also known as peer-to-peer file sharing networks).

the two categories.
\subsection{\label{sec:relatedWork-generalIoT} General Architectures for IoT Patching}
Prior to the Internet of Things (IoT) era, traditional host-centric IT solutions focused on delivering security updates (i.e., patches) by self-hosting the binary updates to make them retrievable and available to clients. 
The availability and reliability of such solutions may require a complex server infrastructure and experienced IT personnel as the number of clients grow, thus making it very expensive for software providers.
Several works targeted the availability and reliability challenges, including  Liu et al.~\cite{liu2015uaas} who suggested a IaaS solution to outsource infrastructure maintenance, and Zhen-hai and Yong-zhi~\cite{zhen2014automatic} who suggested a Maven-based solution to increase effectiveness in case of updates' independence. 
These solutions were not designed towards a large number of deployed devices as in an IoT network.

Following the rapid growth of IoT networks, the work of Yu et al.~\cite{yu2015handling} draws the challenges of securing IoT devices with traditional security, host-centric IT solutions such as anti-virus and software patches. 
Mainly, the diversity, cyber-physical coupling and scale of IoT devices, forces these security IT solutions to a paradigm shift. 
Several works focused on the diversity by suggesting solutions for specific scenarios~\cite{onuma2017ecu}, and the updating in a not-trustworthy setting whereas a device may already be infected~\cite{huth2016secure,kim2017remote}. 
However, the abovementioned works do not address the scalability challenge. 

Lee and Lee~\cite{lee2017blockchain} proposed a firmware update scheme in an IoT environment based on custom blockchain with a single-purpose block structure, combined with a peer-to-peer file sharing network as BitTorrent for the distribution of updates, to enhance availability, integrity and versions traceability of updates.
This suggestion was improved by Boudguiga et al.~\cite{boudguiga2017towards} with the addition of trusted innocuousness checking nodes that are in charge of verifying the patch before it becomes available to deployed IoT devices.\\
Notably, both of the latter solutions acknowledged the naiveness in uploading an entire file to the blockchain, hence delegated the distribution effort to off-chain approaches. 
Nonetheless, they provide no incentive for non-vendor nodes to join the network, and may thus struggle with similar limitations as a centralized vendor network.

Another approach worth mentioning is IOTA~\cite{TheTangle}, a distributed micro-transactions ledger for IoT devices. 
IOTA is designed for the exchange of services among IoT devices (e.g., electricity for cooling), as opposed to IoT devices patching. 
Moreover, IOTA's incentive mechanism is based on the need of a device to issue a service request and is therefore unfit for patch delivery.

In our proposed solution we extend the use of a blockchain-based network for IoT update delivery, and specifically address the issue of an incentive for providing the updates (by non-vendor nodes). 
In addition, the proposed solution eliminates the need for trust between the security update distributor and the consumer (IoT device), thus allowing rapid scale out.

\subsection{\label{sec:relatedWork-p2p} Decentralized Storage Networks}
Peer-to-peer (P2P) file sharing is a technology for the sharing of digital media. 
Imbalanced use (e.g., peers that avoid uploading) can dramatically reduce the scale and availability of the network and therefore, sharing incentivization is necessary.
The first examples of P2P file sharing incentivization mechanisms are the BitTorrent's choking algorithm~\cite{cohen2003incentives} and Gnutella's free riding prevention~\cite{hughes2005free}, in which the download rate for a user that downloads content yet avoids uploading content, is penalized.
Though these mechanisms have been proven useful for symmetric users that wish to both download and upload content~\cite{legout2006rarest}, they are not designed for an asymmetric sharing scheme. 

The asymmetric file sharing scheme is composed of two types of end-users: distributors that host files and consumers that wish to acquire them. 
Proper incentivization of distributors within this scheme can be more challenging than in the symmetric case. 
However, with the emergence of the blockchain technology that allows data and currency exchange in the absence of trust among its users, new forms of incentivization are unveiled. 
The most common incentive is digital currency compensation for either providing storage or bandwidth, which is the case for blockchain-based decentralized storage solutions such as Swarm~\footnote{\url{https://github.com/ethersphere/swarm}}, Filecoin~\footnote{\url{https://filecoin.io/filecoin.pdf}}, Storj~\cite{wilkinson2014storj}, and Siacoin~\cite{vorick2014sia}.

Regardless of these general incentivization schemes, software patches distribution is a special setting that remains unaddressed, in which there should be no more than a single compensation per patch delivered to a device.
This setting, which is the essence of an IoT software updates distribution system, is the main novelty of this P2P file sharing system.

\section{\label{sec:framework}Proposed Framework}
The proposed framework allows vendors (i.e., manufacturers of IoT devices) a secure and scalable delivery of security updates to their deployed IoT devices. 
It is referred to as a framework because of its general requirements which allow a variety of concrete implementations. The framework is comprised of two networks. 
The first is a decentralized storage network (DSN) (see Section ~\ref{subsec:dsn}), in which security updates are transferred between network nodes, and the second is a blockchain network (see Section ~\ref{subsec:blockchain-network}), in which vendors commit to payment in exchange for delivery in the form of smart contracts, thus incentivizing delivery within the DSN. 
Both of these networks are facilitated by the three types of nodes participating in the IoT updates delivery network (see Section ~\ref{subsec:nodes}): vendors, IoT devices and distributors. 
These nodes participate in both networks and interact with each other to support the paid-for delivery of updates.

\begin{figure*}[ht]
  \centering
  \includegraphics[width=11cm,height=9cm]{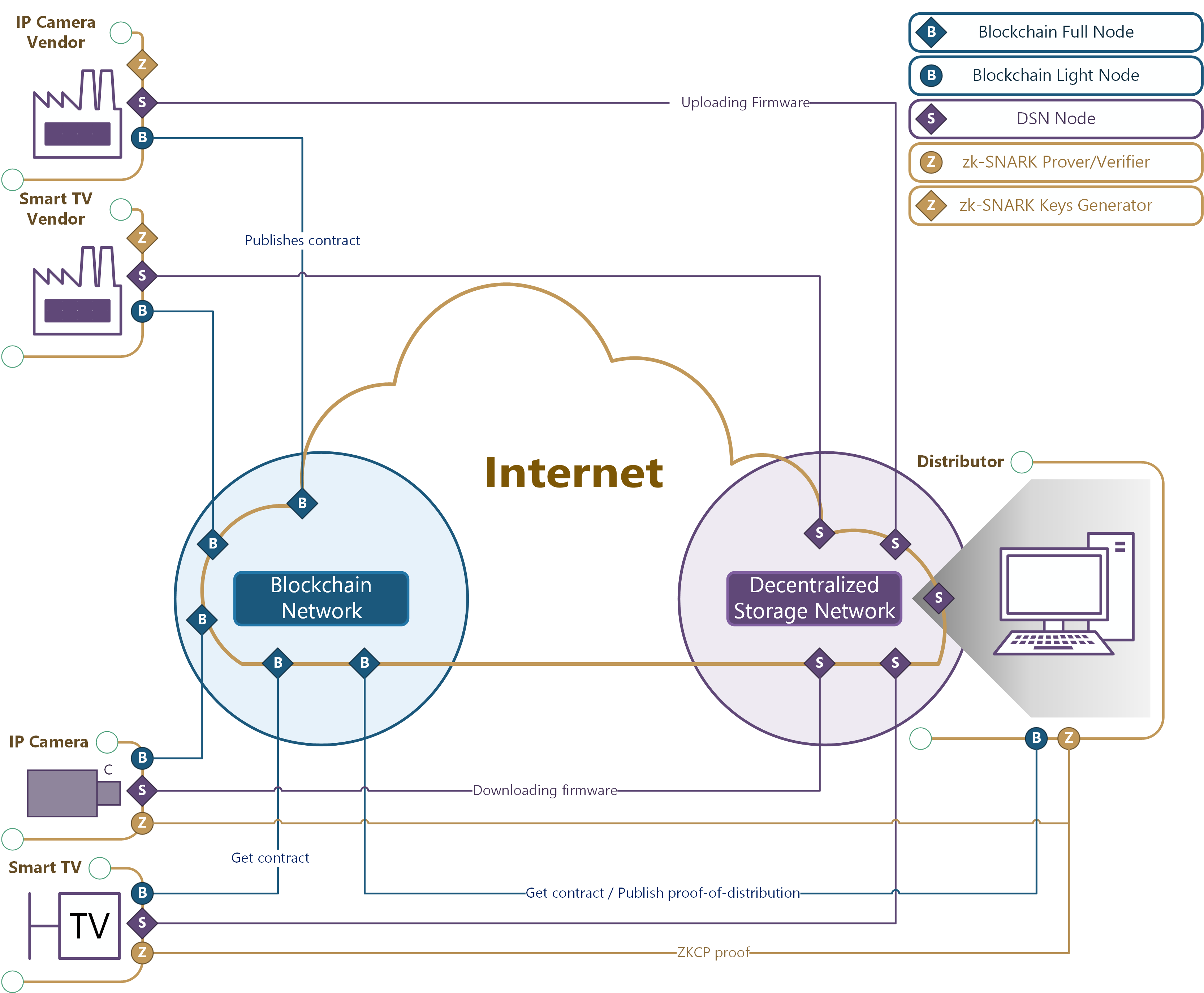}
  \caption{The architecture of the proposed framework.}
  \label{fig:arch}
\end{figure*}

The delivery process (portrayed in Figure~\ref{fig:arch}) is triggered when a vendor releases a new security update that should be delivered to its deployed devices. 
The vendor delegates the task of serving security updates to its deployed IoT devices, by publishing a smart contract in the blockchain network in which he commits to providing coins in exchange for a \textit{proof-of-distribution}, i.e., providing a proof of delivering the update to a deployed device (see Section ~\ref{subsec:contract}). 
Also, the vendor hosts the new security update, enabling other nodes to consume via the DSN, for a limited and short period of time.\\
Upon publication of the smart contract, distributors can engage by downloading the update from the DSN, thus making it available for the IoT devices.\\ 
The IoT devices become aware of an update release with the publication of the smart contract and can now consume the update from distributors over the DSN (see Section ~\ref{subsec:distribution}).
The update exchange between a deployed device and a distributor relies on a trustless swap in which the deployed device is updated and the distributor acquires a proof-of-distribution (see Section ~\ref{subsec:exchange}), thus paid by the smart contract (see Section ~\ref{subsec:proofpub}).

\subsection{\label{subsec:dsn}Decentralized Storage Network (DSN)}
The decentralized storage network (DSN) must be accessible by all nodes (peers) in the network. 
Accessibility can be achieved with a trackerless peer discovery scheme, using a distributed hash table (DHT). 
This suggested scheme is similar to that of BitTorrent~\cite{cohen2003incentives} and IPFS~\cite{benet2014ipfs}.

The connected DSN nodes can both consume and serve files. 
The set of DSN nodes is denoted as follows: 
\[
S = \{s_1, \dots, s_{n_S}\}
\]

\subsection{\label{subsec:blockchain-network}Blockchain Network}
The blockchain network must meet the following properties:
\begin{enumerate}
\item Permissionless - any interested party (e.g., an anonymous person, organization, or company) can read and post messages on the blockchain network.
\item Support an intrinsic digital currency – i.e., the blockchain can be interpreted as a public ledger in which each party has a currency balance.
\item Support smart contracts (as described in Section~\ref{sec:background-smartContracts}).
\end{enumerate}

At the time of this writing, Ethereum~\cite{buterin2014next} is the most widely adopted example of such an open permissionless blockchain which also supports Turing-complete smart contract languages, and thus is a recommended realization of our framework's required blockchain network.

The set of blockchain network nodes is denoted as:
\[
B=\{b_1, \dots, b_{n_B}\}
\]

\subsection{\label{subsec:nodes}IoT Updates Delivery Network}
\subsubsection{\label{subsubsec:vendor-nodes}Vendor nodes} 
Vendor nodes are host machines that are owned by manufacturers of IoT devices. 
All vendor nodes must participate in both the DSN and the blockchain network. 
Within the blockchain network, vendor nodes must be able to function as both a wallet and a network routing node. 
The set of vendor nodes is denoted as:
\[
V = \{v_1, \dots, v_m\}
\] where
\[
V \subset S \wedge V \subset B
\]

Every vendor node, $v_i$, must maintain the following:
\begin{enumerate}
\item a master secret/public keys pair $(sk^{v_i}, pk^{v_i})$.
\item a list of its self-manufactured IoT devices public keys.
Each device must have a unique pair of keys. This can be realized by having the vendor "burning" a new secret key into each new manufactured IoT device and adding the key to its list.
\end{enumerate} 

\subsubsection{IoT nodes}
IoT nodes are machines that require updates by their manufacturing vendor. 
All IoT nodes participate in both the DSN and the blockchain network. 
Within the blockchain network, IoT nodes must be able to function as a network routing node.
For every vendor $v_i \in V$, the set of its IoT nodes is denoted:
\[
O_i= \{o_{i1}, \dots, o_{in}\}
\] where
\[
\forall i \in [1, m] : O_i \subset B \wedge O_i \subset S
\]\\
Each IoT object, $o_{ik} \in O_i$, maintains the following:
\begin{enumerate}
\item a secret/public keys pair $(sk^{o_{ik}}, pk^{o_{ik}})$.
\item public key of its manufacturing vendor (i.e., $pk^{v_i}$).\newline
\end{enumerate}

The last prerequisite can also be met by having the vendor "burn" its own public key $pk^{v_i}$ into the IoT device (thereby adding to the suggestion made in Section \ref{subsubsec:vendor-nodes}, we can have the vendor burn the pair $(pk^{v_i}, sk^{o_{ik}})$ into the deployed device).\\
The requirements of IoT nodes can be realized efficiently in disk space and memory, especially given that IoT devices are often limited in these resources.\\
IoT nodes are not required to store a complete copy of the blockchain. 
Instead, they can rely on a trusted available blockchain node (e.g., a gateway), or reduce the trust needed by consisting of a light client such as under development in Ethereum~\footnote{\url{https://github.com/ethereum/wiki/wiki/Light-client-protocol}}. 
The purpose of the light client protocol is to allow light nodes to download only block headers as they appear, while fetching other parts of the blockchain on-demand, and be sure that the data is correct, provided that the majority of miners are following the protocol correctly, and that at least one honest verifying full node exists and serves the accurate blockchain state.\\
An alternative memory efficient design could be supported by Non-Interactive Proofs of Proof-of-Work \cite{kiayias2017non} which could require only a soft-fork to existing blockchain protocols, and enable verification of a specific blockchain property, requiring only resources logarithmic in the length of the blockchain.\\
In addition, having the IoT function as a DSN node also does not impose significant space constraints. The IoT node is not required to share files in the network, and in the cases in which of the distributed hash table is used (e.g., the Kademlia DHT ~\cite{maymounkov2002kademlia} used in BitTorrent) as the peer discovery scheme for DSN, only a small constant amount of memory is required for maintaining internal routing tables.

\subsubsection{Distributor nodes}
Distributor nodes are host machines that participate in an open bid for proofs-of-distribution.
They must participate in both the DSN and the blockchain-based network. 
Within the blockchain network, distributor nodes must be able to function as both a wallet and a network routing node.
Distributor nodes are denoted as:
\[
D = \{d_1, \dots, d_{n_D}\}
\]
where
\[
D \subset S \wedge D \subset B
\]

\subsection{\label{subsec:protocol}Protocol}
\begin{figure*}[ht]
  \includegraphics[width=\paperwidth, trim={0 10cm 0 0}, clip]{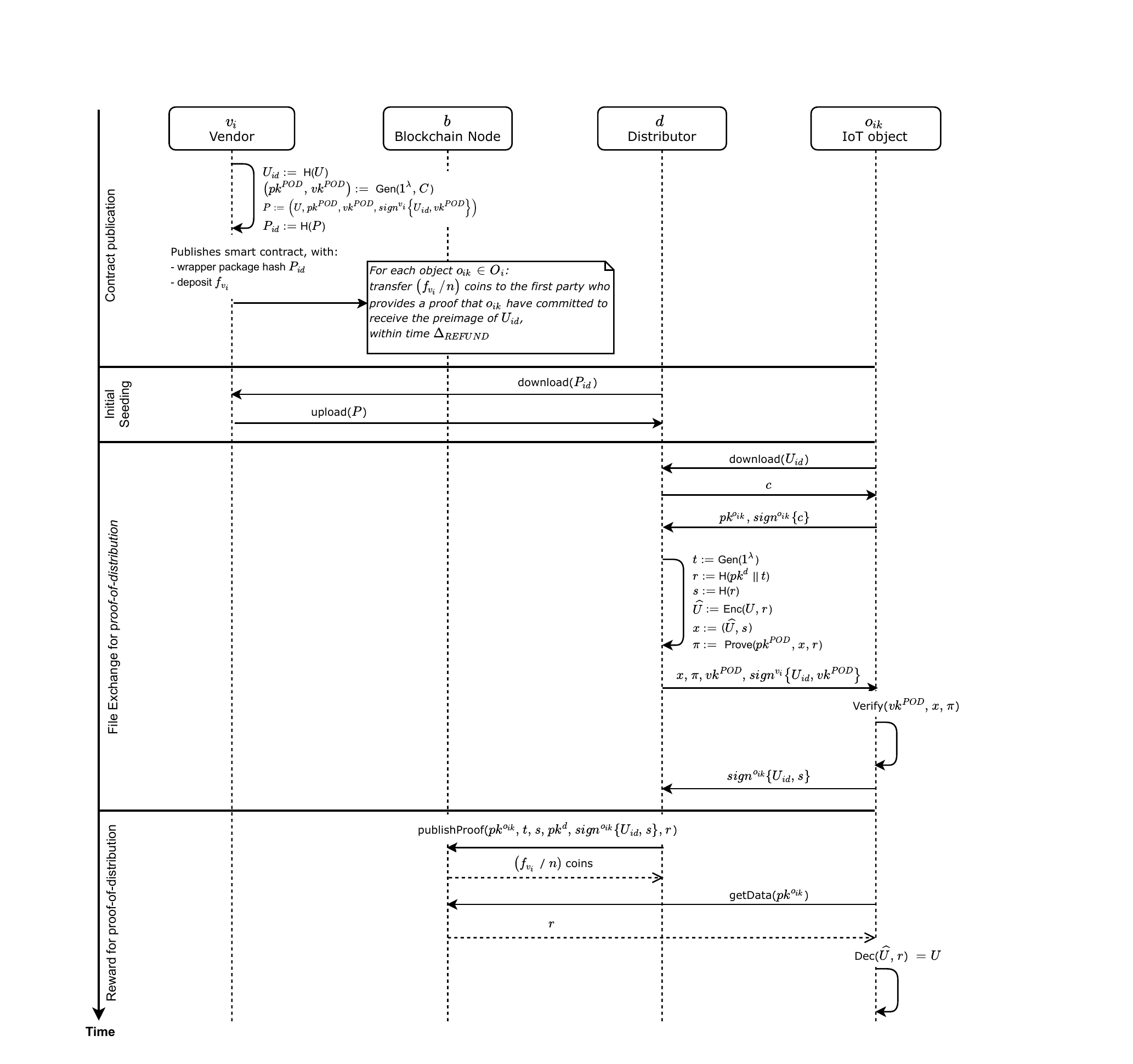}
  \caption{IoT software update protocol sketch. Here $C$ is a decision circuit, $x$ is the instance and $r$ is the secret witness, for the NP statement\newline $\exists r$ s.t. $H(r) = s \land H(Dec(\hat{U}, r)) = U_{id}$}
  \label{fig:sequence-diagram}
\end{figure*}

In this section, we describe the protocol used for a vendor $v_i \in V$ to send an update file $U$ destined to a group of IoT objects manufactured by the vendor, $O_i = \{o_{i1}, \dots, o_{in}\}$ (see Figure~\ref{fig:sequence-diagram} for the protocol sketch).

\subsubsection{\label{subsec:contract} Contract Publication}
Upon releasing a new update $U$, vendor $v_i$ performs the following phases:
\begin{enumerate}

\item Hashes the update file $U_{id} := H(U)$.
\item Generates the zk-SNARKs public proving/verification keys: 
$(pk^{POD}, vk^{POD}) := Gen(1^{\lambda}, C)$.\\
Here $\lambda$ is a security parameter, and $C$ is a decision circuit.
\item Computes update file wrapping package:\\ $P := (U, pk^{POD}, vk^{POD}, sign^{v_i}\{U_{id}, vk^{POD}\})$.
\item Hashes $P_{id} := H(P)$.
\item Sets $\Delta_{REFUND}$ to an acceptable time duration for a patch to be effective.
\item Posts a transaction to the underlying blockchain network which deploys a smart contract (in Algorithm~\ref{alg:contract} we describe pseudo-code for the contract suitable for Solidity), with a deposit $f_{v_i}$ of money, the hash of the update file wrapping package $P_{id}$, and a program which essentially says:\newline
\begin{quote}
For each object $o_{ik} \in O_i$:\\ 
Transfer $(f_{v_i} \setminus n)$ coins to the first party who provides proof that $o_{ik}$ have committed to receive the hash preimage of $U_{id}$, within time $\Delta_{REFUND}$.\newline
\end{quote}

Assigning a time limit $\Delta_{REFUND}$ for publishing proofs is important so that the vendor can collect its refund if some of the IoT objects were not served after some set time (e.g., because sufficient time has passed such that a newer update file must be distributed, or because some remote objects are no longer active).\\
Note that addressing the specific IoT devices for any security update becomes possible as the vendor maintains the list of their public keys.
\end{enumerate}

\IncMargin{1em}
\definecolor{seagreen}{rgb}{0.18, 0.55, 0.34}
\SetAlFnt{\small\color{black}\tt}
\LinesNumbered
\renewcommand{\DataSty}[1]{{\color{yellow}\texttt{#1}}}
\renewcommand{\KwSty}[1]{{\color{blue}\texttt{#1}}}
\renewcommand{\ProgSty}[1]{{\color{black}\texttt{#1}}}
\renewcommand{\FuncSty}[1]{{\color{black}\texttt{#1}}}
\renewcommand{\CommentSty}[1]{{\color{seagreen}\texttt{#1}}}
\renewcommand{\ArgSty}[1]{{\color{black}\texttt{#1}}}
\renewcommand{\FuncArgSty}[1]{{\color{black}\texttt{#1}}}
\renewcommand{\ProcArgSty}[1]{{\color{black}\texttt{#1}}}
\renewcommand{\ProcNameSty}[1]{{\color{black}\texttt{#1}}}
\renewcommand{\NlSty}[1]{{\color{black}\texttt{#1}}}
\begin{algorithm*}
\SetKwProg{Fn}{function}{}{}
\SetKwProg{Contract}{contract}{}{}
\SetKwData{NumOfUpdatedObjects}{numOfUpdatedObjects}
\SetKwIF{If}{ElseIf}{Else}{if}{}{else if}{else}{end if}
\SetKwFunction{ProofsOfDistributionBid}{ProofsOfDistributionBid}
\SetKwFunction{PublishProof}{publishProof}
\SetKwFunction{WithdrawFunds}{withdrawFunds}
\Contract{\ProofsOfDistributionBid}{
  \BlankLine
  \Fn{\ProofsOfDistributionBid{v, e, h, o}}{
      \tcp{Constructor}
      owner $\leftarrow$ v\\
      expiration $\leftarrow$ e\\
      updateHash $\leftarrow$ h\\
      iot\_objects $\leftarrow$ o\\
      numOfUpdatedObjects $\leftarrow$ 0\\
      balance $\leftarrow$ value \tcp{Initial deposit}
  }
  \BlankLine
  \Fn{\PublishProof{pk\_o, t, s, pk\_d, signature, r}}{
      \lIf {block.timestamp $\geq$ expiration}{return}
      \lIf {!iot\_objects[pk\_o] $\lor$ iot\_objects[pk\_o].r $\neq$ null}{return}
      \lIf {r $\neq$ H(pk\_d$||$t)}{return}
      \lIf {s $\neq$ H(r)}{return}
      \lIf {verifySig(pk\_o, signature, updateHash$||$s)}{return}
      iot\_objects[pk\_o].r $\leftarrow$ r\\
      transfer(balance $\div$ (n - numOfUpdatedObjects), pk\_d) \tcp{Decrease balance}
      numOfUpdatedObjects $\leftarrow$ numOfUpdatedObjects + 1\\
      emitEvent("KeyRevealed", pk\_o, r)
  }
  \BlankLine
  \Fn{\WithdrawFunds{}}{
      \lIf {block.timestamp $<$ expiration}{return}
      \lIf {msg.sender $\neq$ owner}{return}
      transfer(balance, owner)
  }
}
\caption{Pseudocode for the \textit{Proofs-of-Distribution} bid contract}\label{alg:contract}
\end{algorithm*}
\DecMargin{1em}

\subsubsection{\label{subsec:distribution} Update File Initial Seeding}

Routinely, distributor nodes (denoted $D$) and the manufactured IoT objects (denoted $O_i$) are watching the blockchain for an indication that a relevant smart contract is deployed by a vendor known by his public key $pk^{v_i}$. This is achievable, for example, by having the distributors and IoT objects to watch a single "factory contract" in which upon receiving a message with the variable arguments, will create a new instance of the child contract (described in Algorithm~\ref{alg:contract}) and trigger an event with the public key of the bidding vendor and the address of the newly created contract.\\
Upon receiving this contract, the following process starts:
\begin{enumerate}
\item Distributors start to request downloading the update file wrapper package corresponding to the hash $P_{id}$ which was published in the smart contract, using the underlying DSN.
\item The vendor $v_i$ transfers (i.e., seeds) the package $P$ via the DSN (off the blockchain). 
The file is then propagated until it eventually reaches some distributor $d \in D$ that wishes to participate in the open bid for the proofs-of-distribution.\\
It should be made clear that this initial seeding by the vendor is meant to last a limited amount of time until a sufficient mass of distributors has downloaded the file, and from that point the distribution is expected to be performed completely by the competing distributors.
\item After completely downloading the package $P$, $d$ verifies that:\\ $VerifySig(pk^{v_i}, sign^{v_i}\{U_{id}, vk^{POD}\}, U_{id} || vk^{POD}) = 1$.
\item $d$ registers itself in the underlying DSN peer discovery scheme, such as a distributed hash table (DHT), as a holder of the file $U$ using its file hash $U_{id}$.
\end{enumerate}

\subsubsection{\label{subsec:exchange} Update File Exchange for a Proof-of-Distribution}
In this procedure, a distributor and an IoT object are conducting a two-party protocol, in which they perform a fair exchange of the update file for a proof-of-distribution, a digital signature given by the IoT object which can reward the distributor with a payment when sent to the smart-contract.\\
The fair exchange is achieved in a trustless manner by applying a zero-knowledge contingent payment (ZKCP~\cite{wiki2011zero}) which makes use of a zero-knowledge proof generated by the distributor and verified by the IoT object. This part is done in a protocol which is external to the blockchain and therefore does not require any specific modifications to the scripting language of the underlying blockchain nor does it add any burden to the blockchain nodes (so it does not cost any digital currency fees or \emph{gas} in the Ethereum jargon).\\
In details, this process consists of the following steps:
\begin{enumerate}
\item $o_{ik} \in O_i$ performs a lookup for the file $U$ using the publicly available file hash $h_U$ from the smart contract deployed by $v_i$, and it reaches the distributor $d$ by using the underlying DSN peer-discovery scheme (e.g., DHT).\\
It then sends $d$ a request to download the update file which is the hash preimage of $U_{id}$.
\item $d$ sends a challenge $c$ to $o_{ik}$ to sign on.
\item $o_{ik}$:\\
- Computes $sign^{o_{ik}}\{c\} := Sign(sk^{o_{ik}}, c)$.\\
- Sends the tuple $(pk^{o_{ik}}, sign^{o_{ik}}\{c\})$ to $d$.
\item $d$:\\
- Verifies that $pk^{o_{ik}} \in O_i$.\\
- Verifies that $VerifySig(pk^{o_{ik}},sign^{o_{ik}}\{c\},c) = 1$.\\
- Computes $t := Gen(1^{\lambda})$, where $Gen$ is some secure random key generator for the security parameter $\lambda$.\\
- Computes $r := H(pk^d || t)$.\\
- Computes $s := H(r)$.\\
- Computes $\hat{U} := Enc(U, r)$ where $Enc$ is some symmetric encryption algorithm (e.g., AES).\\
- Computes $x := (\hat{U}, s)$.\\
- Computes the zero-knowledge proof:\\
$\pi := Prove(pk^{POD}, x, r)$\\
for the NP statement:\\ $\exists r$ s.t. $H(r) = s \land H(Dec(\hat{U}, r)) = U_{id}$.\\
- Sends the tuple $(x, \pi, vk^{VOD}, sign^{v_i}\{U_{id}, vk^{POD}\})$ to $o_{ik}$.
\item $o_{ik}$:\\
- Verifies that:\\ $VerifySig(pk^{v_i}, sign^{v_i}\{U_{id}, vk^{POD}\}, U_{id} || vk^{POD}) = 1$.\\
- Verifies that $Verify(vk^{POD}, x, \pi) = 1$.\\
- Sends the tuple $(sign^{o_{ik}}\{U_{id}, s\})$ to $d$.
\end{enumerate}

\subsubsection{\label{subsec:proofpub} Reward Claim for a Proof-of-Distribution}
As a distributor $d$ acquires the proof-of-distribution, he collects his reward as follows:
\begin{enumerate}
\item $d$:\\
- Verifies that $VerifySig(pk^{o_{ik}}, sign^{o_{ik}}\{U_{id}, s\}, U_{id}||s) = 1$.\\
- Posts a redeem transaction to the smart contract containing: $(pk^{o_{ik}}, t, s, pk^d, sign^{o_{ik}}\{U_{id}, s\}, r)$ which would release $(f_{v_i} / n)$ coins to $pk^d$ in return, according to the contract description (see function \texttt{publishProof} at Algorithm \ref{alg:contract}). 
\item $o_{ik}$:\\
- Watches the blockchain nodes for a relevant event which is associated with the device's public key, $pk^{o_{ik}}$, with the associated decrypting key $r$ (see event \texttt{"KeyRevealed"} in Algorithm \ref{alg:contract}).\\
- Computes $Dec(\hat{U}, r)$ to get $U$.
\end{enumerate}

\section{\label{sec:ta}Security}
The proposed framework encourages distributor nodes to participate in the protocol by delivering security updates to IoT devices in exchange for payment compensation. 
Therefore, a secure implementation of the system would include the guarantee for a fair exchange; i.e., regardless of a trust setting, every distributor node that delivers an update to a device is compensated correctly.
This section briefly discusses potential attacks and their handling (Section~\ref{subsec:addition-attack}) and ends with a formal claim and proof of the fair exchange (Section~\ref{subsec:pos-proofs}), thus guaranteeing that the system is secure. 

\subsection{\label{subsec:addition-attack}Threat Analysis}
In this work we consider the Dolev and Yao attacker model \cite{dolev1983security}. 
In this case an attacker can \textit{read}, \textit{send} and \textit{drop} any transaction sent to the blockchain or any other network packet. 
In addition, the attacker can be a passive party eavesdropping on the network packets, or it can be active by injecting, replaying, or filtering any message to anyone of the parties involved.
Based on this attack model we consider the following threats to the system:\\
\textbf{Denial-of-service (DoS).} An attacker can target a vendor by preventing legitimate transactions from appearing on the ledger, thus preventing the vendor from posting a new software update. 
That kind of an attack is typically referred to as a censorship attack. By construction, the attacker will need to control the majority of the mining power of the network, which is presumably difficult.\\
An alternative DoS attack is preventing a vendor from uploading new software update to the distributors network, or preventing an IoT device from downloading new software from the distributor. 
These attacks are mitigated using the multiple independent access points provided by the decentralized storage network.
Using a decentralized peer-discovery scheme, such as a distributed hash table (DHT) as used in BitTorrent \cite{jimenez2009connectivity} and IPFS \cite{benet2014ipfs} as examples, further allows fault tolerance and reduces single points of failure.
Furthermore, an attacker could prevent the connection from the IoT device to the blockchain network, thus preventing the device from being notified about new software updates.
However, the various access points given by the blockchain network assist in mitigating this threat.\\
\textbf{Compromising software integrity.} An attacker can impersonate a vendor and try to publish a malicious software update to the IoT devices.
The integrity of the firmware update is ensured by the ability of the IoT device to verify the received file $U$ by its hash $U_{id}$ specified in the contract. 
The contract was created by the vendor in a transaction signed by him and verified against his known public key.\\
\textbf{Software downgrade attack.} An attacker may cause an IoT device to downgrade its software to an outdated version (which may include known vulnerabilities). 
This threat is mitigated using the consensus regarding the order of the blockchain's transactions, which implies ordering of the updates versions.

\subsection{\label{subsec:pos-proofs}Proof of Fair Exchange}
Let $U_{ij}$ be the $j$th update of the vendor $v_i$.\\
\textbf{Claim:} Except a negligible probability in the security parameter $\lambda$, a distributer node $d_{q}$ can receive the payment compensation $C(U_{ij},o_{ik})$ if and only if:
\begin{enumerate}
  \item $d_{q}$ delivered the security update $U_{ij}$ to the IoT device $o_{ik}$, and
  \item No other distributor $d_{p}$ $(p \neq q)$ received $C(U_{ij},o_{ik})$
\end{enumerate}
\textbf{Proof:} Let $SC_{ij}$ be the smart contract published by the vendor $v_i$ towards the delivery and compensation of the update $U_{ij}$. 
The contract $SC_{ij}$ contains the set of IoT devices $O_{i}$, with their proofs-of-distribution (if received). 
By the contract construction, for every $r_{x}$, a successful proof-of-distribution for the update $U_{ij}$ and device $o_{ix}$, the object $o_{ix}$ is assigned with $r_{x}$ only if it was not previously assigned.\\
For any distributor $d_{p}$ ($p \neq q$) to receive $C(U_{ij},o_{ik})$ prior to $d_{q}$, a published transaction reporting the proof-of-distribution $d_{p}$ is required for $U_{ij},o_{ik}$. 
If this transaction is indeed published, $O_{i}$ already has an $r$ assignment for $o_{ik}$, thus a distributor node $d_{q}$ cannot receive $C(U_{ij},o_{ik})$ as required. 
We therefore, continue by assuming $C(U_{ij},o_{ik})$ was not yet received by any distributor.

Assuming $C(U_{ij},o_{ik})$ is still available, in order for $d_{q}$ to receive it -- the smart contract $SC_{ij}$ must induce $verify(\cdot) = 1$. 
Therefore, if $d_{q}$ did not deliver the security update $U_{ij}$ to the IoT device $o_{ik}$ he is required to provide the contract with a forged published proof and yet suffice $verify(\cdot) = 1$. 
Based on the security parameter of the system $\lambda$, a distributor $d_{q}$ can forge this proof only if:
\begin{enumerate}
\item The distributor forges $o_{ik}$ digital signature; i.e., $d_{q}$ forges $sign^o_{ik}\{U_{id}, s, pk^d\}$.
\item The distributor forges the entire zk-SNARKs' proof-of-knowledge.
\item The distributor uses the proof-of-distribution of another distributor, $d_{q'}$, as his own. i.e., $d_{q}$ manages to find a hash preimage $t'$ s.t $H(pk^{d_q}||t')=H(pk^{d_{q'}}||t)=r$.
\end{enumerate}
Based on their definitions and the use of the $\lambda$ parameter in the protocol, digital signatures, zk-SNARKs and hash preimaging are unforgeable except with negligible probability in $\lambda$.
Therefore, $d_{q}$ cannot receive $C(U_{ij},o_{ik})$ without a proof-of-distribution except with a negligible probability in $\lambda$.

In the other direction, by the completeness of zk-SNARKs, a distributor $d_{q}$ that delivers $U_{ij}$ to the IoT device $o_{ik}$ will be able to acquire a proof-of-distribution and can publish it to collect $C(U_{ij},o_{ik})$, assuming that the IoT object will not disconnect in the middle of their two-party protocol. 
If no other distributor collected $C(U_{ij},o_{ik})$, based on the irreversibility of the blockchain, the smart contract is binding, thus $d_{q}$ will be able to receive the $C(U_{ij},o_{ik})$ payment as required.\\
It should be noted that the vendor $v_i$, which has performed the zk-SNARKs setup to generate the proving and verification keys, can theoretically forge proofs to falsely convince IoT devices that he has delivered them the update file. However, since this will serve him only in getting his own vested funds from the smart contract, and beat his own original purpose to update his manufactured devices, we argue that this is not a real limitation of the protocol.

\section{\label{sec:discussion}Discussion}
Implementation of the proposed framework is based on several assumptions. 

\textbf{Amount of data needed to be stored and sent to the blockchain.} Given the need for maximum security from forging signatures on behalf of IoT objects, we are required to actually list the entire set of public keys of the destination IoT objects. In addition, based on the hope that all of the micropayments offered will be redeemed, a similar number of transactions, each containing its corresponding proof-of-distribution along with the public keys of the distributor and the IoT object, would be sent to the blockchain.\\
Some optimizations can be made:
\begin{enumerate}
\item Using batched transactions for sending multiple redeem proofs - constructing the contract so multiple redeem transactions can be aggregated into one. This can be even more useful when combining with a signature aggregation scheme (such as Schnorr signatures) within the smart contract, as is planned to be implemented with Bitcoin~\cite{cryptoeprint:2018:068}. The latter would help reduce the number of signatures needed in cases in which a distributor can serially send aggregated signatures to multiple IoT objects (the multiple messages still must be included in the transaction).
\item Sending an index of the public key in the redeem transaction instead of the whole public key.
\end{enumerate}
At the time of this writing, Ethereum blockchain at its current capacity is able to process 10-15 transactions per second with blocks roughly every 15 seconds, where the median transaction fee is around 25 cents.
As the underlying blockchain is currently the bottleneck of the suggested framework's scalability, other consensus layer scalability solutions~\cite{ScalingEthereum} and platforms can directly benefit our proposed solution.

\textbf{Some trust assumptions regarding IoT objects.} The proposed scheme allows the possibility of an IoT object $o$ to cheat and pretend it is a distributor in order to reduce the need to upload an entire update file $U$ to the blockchain.\\
This can be done by posting a transaction to the smart contract containing the distributor's signature on some string $r$ and its corresponding hash $s$, where $r$ is not an actual random key that decrypts an encryption version of the update file $U$ but is rather just some arbitrary data used to give $o$ the desired micropayment.\\
Even though this dishonest behavior is possible in cases in which the IoT object's client software is hacked or compromised, we argue that the IoT software update micropayment cost was actually encompassed in the cost of the device when it was purchased, and we further argue that the act of updating the software is of greater benefit to the IoT buyer than settling for the coins instead.

\section{\label{sec:conclusion-future-work}Conclusions and Future Work}
We have described a decentralized architecture which utilizes a trustless network to provide software updates to IoT devices. 
It is based on an economic model where compensation and cost are aligned with the requested service.
Service providers (the distributors) have the incentive to perform honestly and quickly, since they know that they will get paid if and only if they will be the first to distribute updates to destination IoT objects, and will earn in proportion to the number of clients served. 
The presented system allows high availability of software updates with fault tolerance, integrity and the ability for a high quality of service in comparison to traditional architectures. 
Also, it is compatible but not limited to existing blockchain platforms such as Ethereum.
For future work, we are interested in investigating performance benchmarks and scalability improvements which will enable the suggested ecosystem to stretch its ability to support the ever-increasing number of IoT devices. 
 
{\footnotesize{}\bibliographystyle{IEEEtran}
\bibliography{Bibliography}
}{\footnotesize \par}

\end{document}